\documentclass[10pt,twocolumn,twoside]{IEEEtran}
\usepackage{amsmath,graphicx}
\usepackage{amssymb}
\usepackage{graphicx}
\usepackage{graphics}
\usepackage{epstopdf}
\usepackage{amsmath}
\usepackage{algorithmicx}
\usepackage{algorithm}
\usepackage{algpseudocode}
\usepackage{multirow}
\usepackage{bm}
\usepackage{enumerate}

\newcommand{\BigO}[1]{\ensuremath{\operatorname{O}\bigl(#1\bigr)}}

\begin{document}

\title{A Note on Shift Retrieval Problems}

\author{Cristian Rusu\thanks{The author is with the Istituto Italiano di Tecnologia (IIT), Genova, Italy. Contact e-mail address: cristian.rusu@iit.it. Demo source code: https://github.com/cristian-rusu-research/shift-invariance}}

\maketitle

\begin{abstract}
In this note, we discuss the shift retrieval problems, both classical and compressed, and provide connections between them using circulant matrices. We review the properties of circulant matrices necessary for our calculations and then show how shifts can be recovered from a single measurement.
\end{abstract}



\section{Shift retrieval problems}


Consider the square circulant matrices defined as:
\begin{equation}
\begin{aligned}
\mathbf{C} =  \text{circ}(\mathbf{c}) \stackrel{\rm def}{=} & \begin{bmatrix}
c_1 & c_{n} & \dots & c_3 & c_2 \\
c_2 & c_1 & \dots & c_4 & c_3 \\
\vdots & \ddots & \ddots & \ddots & \vdots \\
c_{n-1} & c_{n-2} & \dots & c_1 & c_n \\
c_n & c_{n-1} & \dots & c_2 & c_1
\end{bmatrix} \\
= & \begin{bmatrix} \mathbf{c} & \mathbf{P}\mathbf{c} & \mathbf{P}^2\mathbf{c} & \dots & \mathbf{P}^{n-1}\mathbf{c} \end{bmatrix} \in \mathbb{R}^{n \times n}.
\end{aligned}
\label{eq:circulant}
\end{equation}

The matrix $\mathbf{P} \in \mathbb{R}^{n \times n}$ denotes the orthonormal circulant matrix that circularly shifts a target vector $\mathbf{c}$ by one position, i.e., $\mathbf{P} = \text{circ}(\mathbf{e}_2)$ where $\mathbf{e}_2$ is the second vector of the standard basis of $\mathbb{R}^{n}$. Notice that $\mathbf{P}^{q-1} = \text{circ}(\mathbf{e}_q)$ is also orthonormal circulant and denotes a cyclic shift by $q-1$. The eigenvalue factorization of circulant matrices reads:
\begin{equation}
\mathbf{C} = \mathbf{F}^H \mathbf{\Sigma} \mathbf{F},\ \mathbf{\Sigma} = \text{diag}(\mathbf{\sigma}) \in \mathbb{C}^{n \times n},
\label{eq:cfactorizationr}
\end{equation}
where $\mathbf{F} \in \mathbb{C}^{n \times n}$ is the unitary Fourier matrix ($\mathbf{F}^H\mathbf{F} = \mathbf{F}\mathbf{F}^H = \mathbf{I}$) and the diagonal $\mathbf{\sigma} = \sqrt{n}\mathbf{Fc},\ \mathbf{\sigma} \in \mathbb{C}^n$.

The multiplication with $\mathbf{F}$ is equivalent to the application of the Fast Fourier Transform, i.e., $\mathbf{Fc} = \text{FFT}(\mathbf{c})$, while the multiplication with $\mathbf{F}^H$ is equivalent to the inverse Fourier transform, i.e., $\mathbf{F}^H\mathbf{c} =  \text{IFFT}(\mathbf{c})$. Both transforms are applied in $\BigO{n \log n}$ time and memory. 

Given two real-valued matrices $\mathbf{X}$ and $\mathbf{Y}$, both $n \times N$, an immediate application \cite{CDLA, OnLearningRusu} of the eigenvalue factorization with the Fourier matrix is to the solution of the problem:
\begin{equation}
	\underset{\mathbf{\sigma}}{\text{minimize}} \ \|  \mathbf{Y} - \mathbf{CX} \|_F^2,
	\label{eq:minimize}
\end{equation}
whose solution is given by
\begin{equation}
\sigma_1 = \frac{ \mathbf{\tilde{x}}_1^H \mathbf{\tilde{y}}_1 }{ \|\mathbf{\tilde{x}}_1\|_2^2 },\sigma_k = \frac{ \mathbf{\tilde{x}}_k^H \mathbf{\tilde{y}}_k }{ \|\mathbf{\tilde{x}}_k\|_2^2 },\ \sigma_{n-k+2} = \sigma_k^*,k=2,\dots,n,
\label{eq:optimalsolution}
\end{equation}
where $\mathbf{\tilde{y}}_k^T$ and $\mathbf{\tilde{x}}_k^T$ are the rows of $\mathbf{\tilde{Y}} = \mathbf{FY}$ and $\mathbf{\tilde{X}} = \mathbf{FX}$.

\subsection{Classic shift retrieval}

Given two signals $\mathbf{x}, \mathbf{y} \in \mathbb{R}^{n}$ assuming that $\mathbf{y}$ is a cyclic shift of $\mathbf{x}$ in order to find the shift amount we maximize the inner product:
\begin{equation}
\underset{q}{\arg \max} \ |\mathbf{x}^T \mathbf{P}^q \mathbf{y}|,
\label{eq:normalShift}
\end{equation}
where $\mathbf{P}^{q} \in \mathbb{R}^{n \times n}$ denotes a cyclic shift by $q$. The calculations above explicitly find all inner products between $\mathbf{x}$ and all possible circular shifts of $\mathbf{y}$, or vice-versa. Practically, to recover the shift we use the circular cross-correlation theorem:
\begin{equation}
\arg \max \ |\text{IFFT} ( \text{FFT}(\mathbf{x})^* \odot \text{FFT}(\mathbf{y}))|.
\label{eq:cross-correlation}
\end{equation}

The result follows directly from the factorization \eqref{eq:cfactorizationr} by computing correlations between all cyclic shifts of	$\mathbf{x}$ and the vector $\mathbf{y}$ as
\begin{equation}
	\begin{aligned}
 \mathbf{C}^T \mathbf{y} = & \text{circ}(\mathbf{x})^T \mathbf{y}\\
 =&\mathbf{F}^H \text{diag}(\sqrt{n}\mathbf{Fx})^H \mathbf{F} \mathbf{y} \\
	= &  \text{IFFT}(\text{FFT}(\mathbf{x})^* \odot \text{FFT}(\mathbf{y})).
	\end{aligned}
\label{eq:quantitytomaximize}
\end{equation}
Therefore, the problem in \eqref{eq:cross-correlation} becomes that of maximizing $\|  \mathbf{C}^T \mathbf{y}   \|_\infty$. When the signals are complex-values the real part of the objective is considered. In our case, the absolute value removes the distinction between positive and negative correlation of real-valued signals. Equivalently, we could be squaring the quantity to achieve an equivalent effect. In the following section we will make use of the $\ell_2$ norm. Note that if the two signals are circularly shifted version of each other (by the amount $q-1$) then the result of this calculation is $\| \mathbf{x} \|_2^2 = \| \mathbf{y} \|_2^2$ at index $q$.

\subsection{A different perspective on the shift retrieval problem}

In this section, we provide a different view on the classic shift retrieval problem and give the following main result.

\noindent \textbf{Result 1.} We are given two signals $\mathbf{x}$ and $\mathbf{y}$ such that there is a circular shift $q$ between them, i.e., $\mathbf{y} = \mathbf{P}^{q-1} \mathbf{x}$. Then:
\begin{equation}
\text{IFFT} (\text{FFT}(\mathbf{y}) \oslash \text{FFT} (\mathbf{x}) ) = \mathbf{e}_q.
\label{eq:theorem1}
\end{equation}

\noindent \textit{Proof.} Assuming that $\mathbf{y} = \mathbf{P}^{q-1} \mathbf{x}$, and then $\mathbf{y} = \mathbf{P}^{q-1-n}\mathbf{x}$, with $\mathbf{P} = \text{circ}(\mathbf{e}_2)$. We consider the problem
\begin{equation}
	\underset{q}{\text{minimize}} \  \| \mathbf{y} - \mathbf{P}^{q-1} \mathbf{x} \|_F^2.
	\label{eq:minimizePq}
\end{equation}
Use $\mathbf{P}^{q-1} = \mathbf{F}^H \mathbf{\Sigma} \mathbf{F}$ and to develop $\| \mathbf{y} - \mathbf{P}^{q-1} \mathbf{x} \|_F^2
=\|\mathbf{y} - \mathbf{F}^H \mathbf{\Sigma} \mathbf{Fx} \|_F^2
= \| \mathbf{Fy} -  \mathbf{\Sigma} \mathbf{Fx} \|_F^2 
=  \| \mathbf{\tilde{y}} - \mathbf{\Sigma} \mathbf{\tilde{x}} \|_F^2$,
where $\mathbf{\Sigma} = \text{diag}(\mathbf{\sigma}), \mathbf{\sigma} = \mathbf{F e}_q$ (the $q^\text{th}$ column of the Fourier matrix). If we relax the constraint and allow $\mathbf{P}^{q-1}$ to be any circulant matrix, to minimize the Frobenius norm, as the special case of \eqref{eq:minimize} for $N=1$, we have $\sigma_i = \tilde{y}_i / \tilde{x}_i, \ \tilde{x}_i \neq 0,$ and $\mathbf{Fe}_q = \mathbf{\tilde{y}} \oslash \mathbf{\tilde{x}}$.

The assumption that $\tilde{x}_i \neq 0$ seems restrictive (and is missing in \eqref{eq:cross-correlation}). We do not need to apply the inverse Fourier transform but instead compute only $\sigma_i$ where $\tilde{x}_i \neq 0$ and by inspection of all columns of $\mathbf{F}$ on the rows where this quantity was computed we find the shift $q$. Notice that $\sigma_1$ and $\sigma_{\frac{n}{2}+1}$ when $n$ is even are $\{\pm 1\}$ for all columns $j$ and thus they cannot provide an unambiguous answer, on the other hand, in the best case scenario, we would need to compute a single $\sigma_i$. This reduces the complexity of the shift retrieval problem to $\BigO{n}$, as also observed for the compressive shift retrieval result \cite{CSR} which is discussed in the next subsection.  Generally, when $\mathbf{y} = \alpha \mathbf{P}^{q-1} \mathbf{x} + \beta \mathbf{1}$ where $\alpha, \beta \in \mathbb{R}$ then $\text{IFFT} (\mathbf{\tilde{y}} \oslash \mathbf{\tilde{x}}) \! =  \! \alpha \mathbf{e}_q + \left( \beta / \sum_i x_i  \right) \mathbf{1}$, $\mathbf{1} \in \mathbb{R}^n$ is the ones vector.$\hfill \blacksquare$

\noindent \textbf{Remark 1 (Connection to the classical circular cross-correlation theorem).} We can rewrite \eqref{eq:theorem1} as:
\begin{equation}
\begin{aligned}
\text{IFFT} ( \text{FFT}(\mathbf{y}) & \oslash \text{FFT}(\mathbf{x}) ) = \text{IFFT} ( \text{FFT}(\mathbf{y}) \oslash \text{FFT}(\mathbf{x}) ) & \\
= & \text{IFFT} (  \text{FFT}(\mathbf{x})^* \odot \text{FFT}(\mathbf{y}) \oslash | \text{FFT}( \mathbf{x}) |^{ 2} ),
\end{aligned}
\label{eq:similarity}
\end{equation}
where $|\text{FFT}(\mathbf{x})|^{2}$ computes the square absolute values of each element of the Fourier transform of $\mathbf{x}$. Notice that \eqref{eq:theorem1} represents a weighted variant of \eqref{eq:cross-correlation}. If the two signals $\mathbf{x}$ and $\mathbf{y}$ are shifted versions of each other then \eqref{eq:cross-correlation} and \eqref{eq:theorem1} provide the same answer. If this is not the case, or the signals are noisy, then \eqref{eq:theorem1} seems a weaker result in general since the minimizer $\mathbf{P}^{q-1}$ in \eqref{eq:minimizePq} might no longer have $\mathbf{P} = \text{circ}(\mathbf{e}_2)$, but some other circulant matrix that minimizes \eqref{eq:minimizePq}. In this high noise case we might not be able to interpret that the signals are shifted versions of each other. The circulant cross-correlation theorem does not have this feature as it will always provide the maximum correlation between the signals.$\hfill \blacksquare$




\noindent \textbf{Remark 2 (Calculation of the circular shift from one measurement).} Notice that \eqref{eq:theorem1} is equivalent to:
\begin{equation}
	\text{FFT}(\mathbf{y}) \oslash \text{FFT} (\mathbf{x})  = \mathbf{f}_q,
\end{equation}
where $\mathbf{f}_q$ is the $q^\text{th}$ column of the Fourier matrix $\mathbf{F}$. We can find the shift by computing a single entry $\tilde{y}_i/\tilde{x}_i$ and then inspecting the entries of only the $i^\text{th}$ column of the Fourier matrix. $\hfill \blacksquare$

We note that the approach to maximize the quadratic form \eqref{eq:normalShift} and that of norm minimization \eqref{eq:minimizePq} are equivalent since
\begin{equation*}
	\begin{aligned}
	 \| & \mathbf{y} - \mathbf{P}^{q-1} \mathbf{x} \|_F^2 =  \| \mathbf{y} \|_2^2 + \| \mathbf{x} \|_2^2 - 2\mathbf{y}^T \mathbf{P}^{q-1} \mathbf{x}  \text{ and also } \\
	\| & \mathbf{y} - \mathbf{P}^{q-1} \mathbf{x} \|_F^2 = \| \mathbf{Fy} -  \sqrt{n}(\text{diag}(\mathbf{F e}_2))^{q-1} \mathbf{Fx} \|_F^2 \\
	= & \| \mathbf{\tilde{y}} -  \sqrt{n}\text{diag}(\mathbf{F e}_q )\mathbf{\tilde{x}} \|_F^2 \\
	= & \| \mathbf{\tilde{y}} - \sqrt{n} \text{diag}(\mathbf{f}_q )\mathbf{\tilde{x}} \|_F^2\\
	= & \| \mathbf{\tilde{y}} \|_2^2 + \| \mathbf{\tilde{x}} \|_2^2 - 2\sqrt{n}\Re(\mathbf{\tilde{y}}^H\text{diag}(\mathbf{f}_q )\mathbf{\tilde{x}} )\\
	 = & \| \mathbf{\tilde{y}} \|_2^2 + \| \mathbf{\tilde{x}} \|_2^2 - 2\sqrt{n}\sum_{i=1}^n \tilde{y}_i^* \odot \tilde{x}_i \odot f_{iq},
	 \end{aligned}
\end{equation*}
where $f_{iq}$ is an element from the Fourier matrix and the last quantity is real-valued due to the conjugate valued symmetries of the vectors $\mathbf{\tilde{x}}$ and $\mathbf{\tilde{y}}$, and of the columns $\mathbf{f}_q$. The last summation quantity is equivalent to \eqref{eq:cross-correlation} for a fixed $q$. The result \eqref{eq:theorem1} is obtained by allowing the unknown to be the overall general circulant matrix denoted $\mathbf{P}^{q-1}$, not just the power $q$. Finally, note that for real-valued $\mathbf{x}$ and $\mathbf{y}$ we have that $\text{IFFT} ( \text{FFT}(\mathbf{x})^* \odot \text{FFT}(\mathbf{y}))$ is equivalent to $\text{FFT} ( \text{FFT}(\mathbf{x}) \odot \text{FFT}(\mathbf{y})^*)$.$\hfill \blacksquare$

We next look at two more general shift retrieval problems.

\subsection{The 1-to-$N$ shift retrieval problem}

In the previous section, we have assumed that the signals to be compared are singletons (we could call this the 1-to-1 shift retrieval problem). In this section we explore what happens when we want to solve the shift retrieval problem between a signal $\mathbf{x}$ and a group of signals $\mathbf{Y} \in \mathbb{R}^{n \times N}$, i.e., find the shift for the signal $\mathbf{x}$ such that it aligns best with all signals from $\mathbf{Y}$. Just as before, we can approach this problem as maximizing \eqref{eq:quantitytomaximize} or like a minimization problem \eqref{eq:minimizePq}.

In our case, the quantity in \eqref{eq:quantitytomaximize} generalizes to
\begin{equation}
	\underset{q}{\arg \max}\  \| \mathbf{Y}^T \mathbf{P}^{q} \mathbf{x} \|_1,
	\label{eq:quantitytomaximizegeneral}
\end{equation}
and this is equivalent to the approach:
\begin{equation}
\arg \max\ \|\text{circ}(\mathbf{x})^T \mathbf{Y}  \|_\infty,
\end{equation}
 where this is the matrix $\infty$-norm, i.e., $\| \mathbf{Z} \|_\infty = \underset{i}{\max} \sum_j |Z_{ij}| $.  Using \eqref{eq:cfactorizationr}, the quantity $\text{circ}(\mathbf{x})^T \mathbf{Y}$ is equivalent to $\sqrt{n} \text{IFFT} ( \text{diag}(\text{FFT}(\mathbf{x})^*)  \text{FFT}(\mathbf{Y})  )$, whose computational complexity is $O(nN\log n)$.

\noindent \textbf{Result 2.} We are given a signal $\mathbf{x}$ and a group of signals $\mathbf{Y}$, we aim to find the shift that reduces the distance between $\mathbf{x}$ and all the vectors $\mathbf{y}_i$ from $\mathbf{Y}$ in the sense:
\begin{equation}
	\underset{q}{\text{minimize}} \  \| \mathbf{Y} - \mathbf{P}^{q-1} (\mathbf{1}_{1 \times N} \otimes \mathbf{x}) \|_F^2,
	\label{eq:minimizePqforY}
\end{equation}
where $\otimes$ is the Kronecker product. The Frobenius norm quantity is minimized for the $q$ returned by
\begin{equation}
	\arg \max\ \text{FFT} \left( \text{FFT} (\mathbf{x}) \odot \sum_{i=1}^N \text{FFT}( \mathbf{y}_i)^*  \right).
	\label{eq:theorem2}
\end{equation}
\noindent \textit{Proof.} 
We use \eqref{eq:cfactorizationr} and expand the Frobenius quantity:
\begin{equation*}
	\begin{aligned}
	\| & \mathbf{Y} - \mathbf{P}^{q-1}  (\mathbf{1}_{1 \times N} \otimes \mathbf{x}) \|_F^2 = \sum_{i=1}^N \| \mathbf{\tilde{y}}_i - \sqrt{n} \text{diag}(\mathbf{\tilde{x}}) \mathbf{f}_q  \|_F^2 \\
	= & \|  \mathbf{\tilde{Y}} \|_F^2 + N \|  \mathbf{\tilde{x}} \|_2^2 - 2\sqrt{n} \left[\mathbf{\tilde{x}} \odot \left(  \sum_{i=1}^N \mathbf{\tilde{y}}_i^* \right)  \right]^T \mathbf{f}_q.
	\end{aligned}
\end{equation*}
The last quantity is the $q^\text{th}$ element of the Fourier transform of the vector in the square brakets.$\hfill \blacksquare$

\subsection{The $N$-to-$N$ shift retrieval problem}

Finally, in the most general case, we are given two sets of signals $\mathbf{X} \in \mathbb{R}^{n \times N}$ and $\mathbf{Y} \in \mathbb{R}^{n \times N}$ the problem is to find a single shift such that each signal $\mathbf{x}_i$ aligns as best as possible with the corresponding signal $\mathbf{y}_i$. This can be seen as the generalization of the problem in the previous sections.

In this case, the quantity in \eqref{eq:quantitytomaximizegeneral} further generalizes to
\begin{equation}
	\underset{q}{\arg \max}\  \text{trace}(| \mathbf{Y}^T \mathbf{P}^{q} \mathbf{X}|),
	\label{eq:quantitytomaximizegeneral2}
\end{equation}

We state the following result.

\noindent \textbf{Result 3.} We are given the signals $\mathbf{X}$ and $\mathbf{Y}$, we aim to find the shift that reduces the distance between all pairwise $\mathbf{x}_i$ and $\mathbf{y}_i$ in the sense:
\begin{equation}
	\underset{q}{\text{minimize}} \  \| \mathbf{Y} - \mathbf{P}^{q-1} \mathbf{X} \|_F^2.
	\label{eq:minimizePqforXY}
\end{equation}
The problem above is solved for the $q$ returned by
\begin{equation}
	\arg \max\ \text{FFT} \left(  \sum_{i=1}^N \text{FFT}( \mathbf{x}_i) \odot \sum_{i=1}^N \text{FFT}( \mathbf{y}_i)^*  \right).
	\label{eq:theorem22}
\end{equation}
\noindent \textit{Proof.} 
We use \eqref{eq:cfactorizationr}, Result 2 and expand the Frobenius quantity:
\begin{equation*}
	\begin{aligned}
		\| & \mathbf{Y} - \mathbf{P}^{q-1}   \mathbf{X} \|_F^2 = \sum_{i=1}^N \| \mathbf{\tilde{y}}_i - \sqrt{n} \text{diag}(\mathbf{\tilde{x}}_i) \mathbf{f}_q  \|_F^2 \\
		= &  \|  \mathbf{\tilde{Y}} \|_F^2 +  \|  \mathbf{\tilde{X}} \|_F^2 - 2\sqrt{n} \left[\left(  \sum_{i=1}^N \mathbf{\tilde{x}}_i \right) \odot \left(  \sum_{i=1}^N \mathbf{\tilde{y}}_i^* \right)  \right]^T \mathbf{f}_q.\hfill \blacksquare
	\end{aligned}
\end{equation*}

\subsection{The compressive shift retrieval problem}

Recently, the compressive shift retrieval problem has been introduced \cite{CSR, CSR2}. Define the sensing matrix $\mathbf{A} \in \mathbb{C}^{m \times n}, m \leq n,$ and the compressed measurement signals $\mathbf{z} = \mathbf{Ay} \in \mathbb{C}^{m}$ and $\mathbf{v} = \mathbf{Ax} \in \mathbb{C}^{m}$. Assuming that $\mathbf{y}$ is a cyclic shift of $\mathbf{x}$, the goal is to determine the shift from $\mathbf{z}$ and $\mathbf{v}$. Similarly to \eqref{eq:normalShift}, consider the test (Corollary 2 in \cite{CSR2}):
\begin{equation}
\underset{q}{\text{argmax}}\ \Re \{ \mathbf{z}^H \mathbf{\bar{P}}^q \mathbf{v} \},
\label{eq:argmaxq}
\end{equation}
where $\mathbf{\bar{P}}^q = \mathbf{A}\mathbf{P}^q\mathbf{A}^H$. It has been shown that when $\mathbf{A}$ is taken to be a partial Fourier matrix then (Corollary 4 in \cite{CSR2}):
\begin{equation}
\underset{q \in \{ 0, \dots, n-1\} }{\text{max}}\ \Re \left\{ \sum_{i=1}^m z_i^* v_i e^{\frac{-2 \pi j k_i q}{n}} \right\},
\label{eq:maxR}
\end{equation}
recovers the true shift if there exists $p \in \{1,\dots,m\}$ such that $\tilde{x}_{k_p} \neq 0$ (the $k_p^\text{th}$ coefficient of the Fourier transform of $\mathbf{x}$) and $\{1,\dots,n-1\}\frac{k_p}{n}$ contains no integers. The set $\mathcal{K} = \{ k_i \}_{i=1}^m$ contains the indices of the rows contained in the partial Fourier matrix $\mathbf{A}$. Following \cite[Theorem 1]{CSR2}, we assume that the sensing matrix $\mathbf{A}$ obeys: $\mathbf{A}^H\mathbf{AP}^{q-1} = \mathbf{P}^{q-1}\mathbf{A}^H\mathbf{A}$, $\exists \ \alpha \in \mathbb{R}$ such that $\alpha \mathbf{AA}^H = \mathbf{I}$ and all columns of $\mathbf{A}\text{circ}(\mathbf{x})$ are different so that there is no shift ambiguity in the measurements.

The compressive shift retrieval result is partly based on the fact that $\mathbf{A}^H \mathbf{AP}^{q-1} = \mathbf{P}^{q-1}\mathbf{A}^H\mathbf{A}$. Notice that $\mathbf{A}^H\mathbf{A} = \mathbf{F}^H \mathbf{\Sigma F}$ where the diagonal $\mathbf{\Sigma}$ contains $\{0, 1\}$ with ones on the positions where the rows of the Fourier matrix are selected ($\mathcal{K}$). Notice that $\mathbf{A}^H\mathbf{A}$ is a circulant and thus it commutes with $\mathbf{P}^{q-1}$ -- they have the same eigenspace. Also, given a set $\mathcal{K}$ of indices, we define the operation $(\mathbf{a})_\mathcal{K} = \mathbf{b}$ for vectors $\mathbf{a} \in \mathbb{C}^n, \mathbf{b} \in \mathbb{C}^m, m\leq n,$ as attributing values $\mathbf{b}$ in positions $\mathcal{K}$ of $\mathbf{a}$, leaving the rest $\{1, 2, \dots,n \} \backslash \mathcal{K}$ to zero. 




\noindent\textbf{Result 4 (Circulant compressive shift retrieval with a proof based on circulant matrices).} Given $\mathbf{z} = \mathbf{Ay}$ and $\mathbf{v} = \mathbf{Ax}$ where $\mathbf{y} = \mathbf{P}^{q-1}\mathbf{x}$, assuming $v_i \neq 0, i=1,\dots,m$ then:
\begin{equation}
(\mathbf{Fe}_q)_\mathcal{K} = \mathbf{z} \oslash \mathbf{v}.
\label{eq:theorem3}
\end{equation}

\noindent\textit{Proof.} We start again from the least squares problem:
\begin{equation}
\underset{q}{\text{minimize}} \  \| \mathbf{z} - \mathbf{A}\mathbf{P}^{q-1} \mathbf{A}^H \mathbf{v} \| _F^2.
\label{eq:minimizeCCompressive}
\end{equation}
With the assumption that $\mathbf{y} - \mathbf{P}^{q-1} \mathbf{x} = \mathbf{0}$ the objective reaches the zero minimum when $\mathbf{P}^{q-1} = \mathbf{F}^H \mathbf{\Sigma} \mathbf{F}, \mathbf{\Sigma} = \text{diag}(\mathbf{Fe}_q)$: $\mathbf{Ay} - \mathbf{AP}^{q-1} \mathbf{A}^H\mathbf{Ax} =  \mathbf{A} (\mathbf{y} - \mathbf{P}^{q-1}\mathbf{x})$, where we used the commutativity of circulant matrices and that  $\mathbf{A}\mathbf{A}^H = \mathbf{I}$. To develop \eqref{eq:minimizeCCompressive}, start again from \eqref{eq:cfactorizationr} and the expression of the matrix multiplication as $\text{vec}(\mathbf{A} \mathbf{F}^H \mathbf{\Sigma} \mathbf{F} \mathbf{A}^H \mathbf{v}) = \left( (\mathbf{FA}^H\mathbf{v})^T \otimes (\mathbf{AF}^H) \right) \text{vec}(\mathbf{\Sigma})$. We finally obtain:
\begin{equation*}
\begin{aligned}
\| \mathbf{z} -  \mathbf{A} \mathbf{P}^{q-1} \mathbf{A}^H \mathbf{v} \| _F^2 =&  \| \mathbf{z} - \mathbf{A} \mathbf{F}^H \mathbf{\Sigma} \mathbf{F} \mathbf{A}^H \mathbf{v} \| _F^2 \\
= & \| \text{vec}(\mathbf{z}) - \text{vec}(\mathbf{A} \mathbf{F}^H \mathbf{\Sigma} \mathbf{F} \mathbf{A}^H \mathbf{v}) \| _F^2 \\
= & \| \mathbf{z} - \left( (\mathbf{FA}^H\mathbf{v})^T \otimes (\mathbf{AF}^H) \right) \text{vec}(\mathbf{\Sigma}) \|_F^2 \\
= & \| \mathbf{z} -  \mathbf{VFe}_q \|_F^2,
\end{aligned}
\end{equation*}
where the matrix $\mathbf{V} \in \mathbb{R}^{m \times n}$ contains only the columns of the Kronecker product that match the non-zero elements of the diagonal matrix $\mathbf{\Sigma}$. The matrix contains the elements of $\mathbf{v}$ in positions $(k_i, i)$. The second equality holds because the Frobenius norm is elementwise. It follows that $\mathbf{VFe}_q = \mathbf{z}$ and
\begin{equation*}
\begin{aligned}
(\mathbf{Fe}_q)_\mathcal{K} = & \mathbf{V}^H (\mathbf{V}\mathbf{V}^H)^{-1} \mathbf{z}\\
= & \mathbf{V}^H (\mathbf{z} \oslash | \mathbf{v} |^{2}) \\
= & \mathbf{v}^* \odot \mathbf{z} \oslash | \mathbf{v} |^{2}\\
=& \mathbf{z} \oslash \mathbf{v}.
\end{aligned}
\end{equation*}
The compressive shift retrieval is equivalent to \eqref{eq:theorem1}, the regular shift retrieval, on the set of Fourier components $\mathcal{K}$. This is a unified view of the shift retrieval solutions.$\hfill \blacksquare$

In relation to \eqref{eq:maxR}, we use the circulant structures to reach:
\begin{equation*}
\begin{aligned}
\mathbf{z}^H \mathbf{\bar{P}}^q \mathbf{v} = & \mathbf{z}^H \mathbf{A} \mathbf{F}^H \mathbf{\Sigma} \mathbf{F} \mathbf{A}^H \mathbf{v} \\
= & \text{vec}(\mathbf{z}^H \mathbf{A} \mathbf{F}^H \mathbf{\Sigma} \mathbf{F} \mathbf{A}^H \mathbf{v}) \\
= & ( (\mathbf{FA}^H \mathbf{v})^T \otimes (\mathbf{z}^H \mathbf{AF}^H) ) \text{vec}(\mathbf{\Sigma}) \\
= & \mathbf{r} \mathbf{Fe}_{q+1},
\end{aligned}
\end{equation*}
where we expressed the matrix multiplications as a linear transformation on $\mathbf{\Sigma} = \text{diag}(\mathbf{Fe}_{q+1})$, with $q \in \{{0, \dots, n-1}\}$ and $\mathbf{r} \in \mathbb{C}^n$ is the expression in the parenthesis with $(\mathbf{r})_\mathcal{K} = \mathbf{z}^* \odot \mathbf{v}.$ The matrix $\mathbf{FA}^H \in \mathbb{R}^{n \times m}$ is a partial permutation matrix -- only positions $(k_i, i)$ are non-zero. The products with $\mathbf{v}$ and $\mathbf{z}$ produce extended vectors $(\mathbf{v})_\mathcal{K}, (\mathbf{z})_\mathcal{K} \in \mathbb{C}^n$. Thus, maximizing $\mathbf{z}^H \mathbf{\bar{P}}^q \mathbf{v}$ reduces to the selection of $\mathbf{e}_{q+1}$.

Due to the natural appearance of the Fourier matrix $\mathbf{F}$ in the factorization of circulant matrices its rows are also the natural choice in the rows of the measurement matrix $\mathbf{A}$. Cancellations that occur because of this choice lead to the analytic results found. This shows a simple alternative, but equivalent, way to develop the result \eqref{eq:maxR} of \cite{CSR2}.

\section{Conclusion}

In this letter, we provide an overview of several shift retrieval problems, among them how to find the circulant shift between two signals with a single measurement in frequency.

\end{document}